# Lessons learnt from the recent EURADOS intercomparisons in computational dosimetry

Hans Rabus[1,7], Maria Zankl[2,7], José Maria Gómez-Ros[3,7], Carmen Villagrasa[4,7], Jonathan Eakins[5,7], Christelle Huet[4,7], Hrvoje Brkić[6,7], Rick Tanner[5,7]

[1] *Physikalisch-Technische Bundesanstalt (PTB), Abbestrasse 2-12, 10587 Berlin, Germany*
[2] *Helmholtz Zentrum München German Research Center for Environmental Health (HMGU), Neuherberg, Germany*
[3] *Centro de Investigaciones Energéticas, Medioambientales y Tecnológicas (CIEMAT), Madrid, Spain*
[4] *Institut de Radioprotection et de Sûreté Nucléaire (IRSN), Fontenay-aux-Roses, France*
[5] *UK Health Security Agency (UKHSA), Didcot, United Kingdom*
[6] *J. J. Strossmayer University of Osijek (MEFOS), Osijek, Croatia*
[7] *European Radiation Dosimetry Group (EURADOS) e.V, Neuherberg, Germany*

**Abstract**

Organized by Working Group 6 "Computational Dosimetry" of the European Radiation Dosimetry Group (EURADOS), a group of intercomparison exercises was conducted in which participants were asked to solve predefined problems in computational dosimetry. The results of these comparisons were published in a series of articles in this virtual special issue of Radiation Measurements. This paper reviews the experience gained from the various exercises and highlights the resulting conclusions for future exercises, as well as regarding the state of the art and the need for development in terms of quality assurance for computational dosimetry techniques.

## 1. Introduction

The European Radiation Dosimetry Group (EURADOS) is an association of 80 institutions and more than 600 individual members that promotes harmonization and good practice in dosimetry (Rühm et al., 2018, 2020; Harrison et al., 2021). EURADOS has eight working groups dealing with different aspects and application areas of radiation dosimetry. EURADOS Working Group 6 on "Computational Dosimetry" has a cross-cutting character. Its main activities include the organization of comparison exercises (Tanner et al., 2004; Gualdrini et al., 2005; Siebert et al., 2006; Price et al., 2006; Gualdrini et al., 2008; Broggio et al., 2012; Vrba et al., 2014, 2015; Caccia et al., 2017) and training courses (Rabus et al., 2021a) as well as studies on fundamental aspects of computational dosimetry.

Recently, several computational dosimetry exercises have been completed, the results of which are compiled in this virtual special issue of Radiation Measurements (De Saint-Hubert et al., 2021, 2022; Eakins et al., 2021; Gómez-Ros et al., 2021, 2022; Huet et al., 2022; Rabus et al., 2021b; Villagrasa et al., 2022; Zankl et al., 2021a, 2021b, 2021c). This current article is meant as a synopsis and reflection on the common issues found in those different exercises and the lessons learnt on the state of the art in applied computational dosimetry, as well as conclusions for future exercises.

## 2. Overview of the exercises

The exercises can be divided roughly into two classes depending on the nature of their solutions. The first class comprised six exercises on the use of ICRP computational reference phantoms (ICRP, 2009; Zankl et al., 2021b) and one on unfolding of neutron spectra from Bonner sphere measurements (Gómez-Ros et al., 2022). The former will be referred to as "voxel-phantom exercises" throughout this article, the latter as the "Bonner sphere exercise". Despite the quite different nature of the problems to be solved, these exercises had in common that they required the application of well-known methodologies and established computational tools. This allowed the organizers to establish prior reference solutions that could be used to validate the results subsequently submitted by the participants in the exercise.

The second class are exercises where no reference solutions could be established, since one of the objectives of the exercises was evaluating the possible influence of different cross-section models in the codes used by the participants. Two of these exercises were code intercomparisons, one dealing with the calculation of microdosimetric and nanodosimetric quantities (Villagrasa et al., 2019, 2022), and the other with the effects of gold nanoparticles on dose deposition at the microscopic scale (Li et al., 2020a, 2020b; Rabus et al., 2021b, 2021c); the former is called the "uncertainty exercise" in this article and the latter the "nanoparticle exercise". The other two exercises in this class dealt with out-of-field dose calculations. One was about calculating the dose to the foetus during maternal proton therapy treatment and the other was about calculating the secondary neutron fluence. (De Saint-Hubert et al., 2021, 2022). These two exercises are referred to as "foetus dose exercises".

It is important to note that none of the exercises was intended to be a code competition. Rather, the aim was to investigate the dispersion of results when the same problem was solved by different people using different approaches and different codes or the same code with different options. The first class of exercises focused on identifying the state of the art in the application of common methods in computational dosimetry. The second class was more exploratory in nature and aimed to assess the state of the art in terms of the capabilities of codes and approaches. All classes contained tasks of different complexity, and thus different demands on the participants' skills.

The exercises were organized and run by ad-hoc teams composed of EURADOS WG 6 members. In general, the preparation of the exercises involved independent simulations by several team members, with their respective results then cross-referenced to identify potential pitfalls in the proposed exercise definitions and to check whether the tasks were solvable based on the information to be provided. For the exercises with reference solutions, the results of these test simulations were also used to establish those values, (e.g. by taking the mean), as well as gain a handle on the typical levels of uncertainty that may be considered acceptable for them.

In some exercises, templates for reporting results were also provided to the participants.

## 2.1. ICRP reference voxel phantom exercises

Of the voxel-phantom exercises, two involved exposure to an external point source, emitting either $^{60}$Co gamma photons or 10 keV neutrons respectively (Huet et al., 2022). The task to be solved was to calculate the organ absorbed doses and the effective dose for a given exposure duration and activity of the source.

A third voxel-phantom exercise also dealt with point source geometries, but for cases of typical X-ray examinations (Huet et al., 2022). Here the task was more complex, as participants were required to determine the position of the radiation point source in relation to the phantom. In addition, the results were to be presented as conversion coefficients to organ absorbed doses, both from air kerma and kerma area product. This exercise was thus linked to a potential practical application in which the latter quantities are determined as part of the quality assurance of radiological equipment, and the conversion coefficients sought would enable an assessment of the dose absorbed by the patient during the X-ray examination.

The fourth voxel-phantom exercise considered a uniform planar source of 60 keV photons beneath the phantom (mimicking ground contamination by $^{241}$Am) and required participants to calculate organ absorbed dose rates and the effective dose rate for a given area density of the emission rate (Eakins et al., 2021). In the fifth voxel-phantom exercise, a mixed radiation field of gaseous $^{16}$N was considered, emitting beta and high energy gamma radiation from both inside (lung) and outside the human body (Gómez-Ros et al., 2021); the ratios of organ equivalent dose rates to activity concentration were to be determined.

The most extensive voxel-phantom exercise involved an idealized case of internal dosimetry (Zankl et al., 2021c). For the sake of simplicity, hypothetical radionuclides were considered that were uniformly distributed in specified organs and emitted monoenergetic photon or electrons. Here, absorbed fractions and specific absorbed fractions of energy in the "source" organ and in specified "target" organs were to be determined as well as $S$-values for the resulting source and target organ combinations for two specific radionuclides.

*2.2. Bonner sphere spectra unfolding exercise*

The tasks were defined by the counts measured by a set of twelve Bonner spheres of different diameters and known sensitivities (as determined by the organizers with radiation transport simulations), located at a measurement point in one of four known environments: inside the bunker of a medical linac; near a radioactive source; in a simulated workplace field within a neutron calibration facility; or outside a nuclear power plant. The count rates measured by the Bonner spheres were determined by the organizers through Monte Carlo radiation transport simulations of the respective complete measurement setup for each Bonner sphere within its environment. In addition, and to recreate a realistic situation, the count rate from one of the Bonner spheres in one of the scenarios was intentionally given an incorrect value in order to test the participants' ability to detect an erroneous measurement and exclude it when applying the deconvolution procedure (Gómez-Ros et al., 2018, 2022)

*2.3. Micro- and nanodosimetric uncertainty exercise*

In its first phase, this exercise included a microdosimetric and a nanodosimetric intercomparison (Villagrasa et al., 2019). In the frame of the former, frequency distributions of specific energy were to be determined within a microscopic water sphere for different distributions of a low-energy electron emitter with an energy spectrum derived from the internal-conversion Auger emitter $^{125}$I. In the nanodosimetry part, ionization cluster size distributions were to be determined in target spheres of different sizes located at different distances from a point source with the same energy spectrum as in the microdosimetry part. A sensitivity analysis was also performed on the variation of inelastic cross-sections and its consequences for the calculated ionisation cluster size distributions (Villagrasa et al., 2022). In the second phase of the exercise (in preparation), the focus will be on a comparison of the cross-section datasets for low-energy electron transport and a consideration of their impact on the dispersion of nanodosimetric results.

*2.4. Nanoparticle exercise*

In the nanoparticle exercise, the dose enhancement from a gold nanoparticle, as well as the energy spectrum of electrons emitted from it, were to be determined when irradiated with two low-energy X-ray spectra. The geometry was simply a gold sphere in water irradiated with a parallel beam from a plane photon source, the cross-sectional area of which was slightly larger than that of the nanoparticle. Two different nanoparticle diameters were considered, and the dose enhancement was to be determined in spherical water shells around the nanoparticle. (Li et al., 2020a, 2020b; Rabus et al., 2021b, 2021c)

*2.5. Foetal dose during maternal proton therapy*

This exercise consisted of two parts. The first part dealt with the effects of different calculation phantoms for pregnant women, and different code versions of MCNP, on the predicted dose to the foetus during maternal brain proton therapy. The second part dealt with the dependence of the secondary neutron spectra on the Monte Carlo radiation transport codes and nuclear models that were used, and their effects on the calculated and measured neutron doses during proton therapy. (De Saint-Hubert et al., 2021, 2022)

## 3. Experiences from the exercises

*3.1. Observations on participants' results*

In general, the ensemble of participants' results that were submitted initially showed a large scatter. In the exercises for which a reference solution was available, excellent agreement within the expected statistical fluctuations was found in some cases, while others showed significantly larger deviations, which in some individual cases ranged by up to several orders of magnitude. For the tasks without a reference solution, a subset of the reported results also agreed with each other to some extent, while others deviated significantly from this group. In both classes of exercise, the occurrence of extreme outliers was not correlated with the complexity of the problem.

Some of the deviations were attributable to simple errors, such as copy-and-paste mistakes or incorrect arrangement of the results in the given template. Others resulted from misunderstanding how the final results should be normalised (e.g. normalising to the correct quantity but at a different distance from the source than was required). In the voxel-phantom exercise for the case of X-ray examinations, some participants normalized to the value of air kerma free in air at a specific distance from the source or to the entrance surface dose (which includes backscatter) instead of to air kerma free in air at the skin as was requested. In the microdosimetric and nanodosimetric intercomparisons, the normalisation to 'one decay of the electron source' was not always understood by the participants and was also sometimes difficult to implement for some Monte Carlo codes. The use of a logarithmic scale for the microdosimetric quantity (specific energy distribution) also caused problems with proper normalisation.

Many major deviations were caused by the fact that the participants' simulations deviated from the specifications in terms of geometrical dimensions or the quantities to be determined. One example of this was the "nanoparticle" exercise, where only two out of eleven participants implemented the requested geometry correctly, which consisted of a gold sphere irradiated in water by a collimated parallel photon beam of given dimensions. Another example was the voxel phantom exercise on internal dosimetry, where some participants used organ masses that included blood instead of those given in ICRP Publication 110, as was stated in the exercise definition. In some of the voxel-phantom exercises, the choice of the location of the source was also sometimes a problem due to deviation from the correct reference point (e.g. the edge of the phantom array instead of the phantom's skin).

Other causes of major deviations were that some participants were not familiar with certain concepts, such as the normalization quantity "kerma area product" (in the voxel-phantom X-ray exercise) or effective dose; mistakes for the latter included not applying tissue weighting factors correctly, not averaging and summing over the defined set of organs, neglecting to sex-average, or neglecting to apply the correct energy-dependent radiation weighting factor for neutron exposures.

In the voxel-phantom exercises, many participants had problems applying the method recommended for bone marrow dosimetry in (ICRP, 2010). This finding stimulated writing an article to better explain this approach, which is also part of this Special Issue (Zankl et al., 2021a).

As already mentioned, participants whose results differed from the reference solution (class 1) or from the majority of other participants' solutions (class 2) were informed of this fact and asked to revise their solutions. Not all contacted participants responded to this invitation or provided the requested information on details about their simulations.

Of those participants who submitted a revised solution, some did not indicate what they had changed in their computational procedure to arrive at their revised results. This therefore does not give any additional insight into possible similar errors to be expected in future similar exercises, or hints that could have been communicated to the other participants.

In the nanoparticle exercise, where some inconsistencies became evident after the first publication of the results (Li et al., 2020a) and required a thorough re-evaluation (Li et al., 2020b), consistency checks provided clear indications of the causes of the discrepancies for some results. Nevertheless, some of the participants concerned did not provide revised solutions (Rabus et al., 2021b). However, it must be also stressed that the majority of participants were very supportive of the re-analysis of the results and were eager to clarify the origin of the discrepancies found initially.

*3.2. Issues with omitted quality assurance of results*

Many of the anomalies found in the reported data could have been detected by the participants themselves, e.g. through simple plausibility checks of their results. Examples are briefly discussed in the following.

For example, a very general plausibility consideration is that if the irradiation conditions are quite homogeneous, it may be expected that all organ doses will be of broadly similar magnitudes; a single organ dose result differing by several orders of magnitude from the rest of a given participant's dataset ought therefore to be immediately apparent to them as being potentially erroneous. Similarly, if multiple energies are considered, it is unlikely that the value for a single intermediate energy will be entirely outside the range of values for all other energies.

In the Bonner sphere exercise, there were cases of reported results with negative values for the neutron fluence. These physically impossible values, as well as anomalous spectral shapes, could have been identified by simply plotting the results. Some of the reported spectra differed from the reference solutions by several orders of magnitude; such anomalies could have been easily detected if the unfolded spectra had been convolved with the given sensitivities of the Bonner spheres, to verify that the given count rate was then achieved.

In the voxel phantom exercise for internal dosimetry, a simple plausibility check would have been that the absorbed fraction for electrons and low-energy photons must be close to unity in a source organ and quite small for other organs, since these radiations have a short range in condensed matter and therefore deposit their energy close to the point of release. Moreover, some participants in this exercise reported results for absorbed fractions and specific absorbed fractions for which the ratio of these two quantities varied between different energies of the particles emitted from the (monoenergetic) source. However, since this ratio is simply the mass of the organ, it cannot depend on the energy.

In addition, for many of the tasks in the voxel phantom exercises, literature values are available for fairly similar exposure conditions that could have been used for comparison, at least as a first approximation to indicate the expected magnitudes of the results.

When dealing with voxel phantom simulation, one of the simplest checks might be to visualise the problem in order to ensure the proper positioning of the beam, though it is noted that some software packages struggle due to the sizes of these input files. If one is using any variance reduction it should also be ensured that simulations with and without application of these techniques reproduce the same results, albeit with differing statistical uncertainties.

*3.3. Issues with exercise definitions*

In some cases, inadequacies in exercise definitions became apparent while they were already running. In the nanoparticle exercise, for example, one of the quantities to be reported by the participants was the energy spectrum of electrons "in spherical shells" around the nanoparticle, with the radii of the bounding spherical surfaces given.

Most of the participants interpreted this physically undefined quantity as the energy distribution of the electrons entering the respective volume. However, one participant determined the energy distribution of the balance of the number of electrons traversing the surfaces of the respective volume and withdrew her results on the assumption that the observed negative frequencies indicated an error that she could not locate.

Another problem with this part of the nanoparticle exercise was that there was no default energy binning, so participants chose very different values for the bin size, with some using logarithmic binning and others using linear binning. The large statistical variations in the results obtained with small energy bin sizes masked the variations between the different results when plotted together (Li et al., 2020a).

In the exercise on Bonner sphere spectrum unfolding, it was found during the analysis that in one of the scenarios considered, there was an interference of the Bonner sphere response due to backscattering of neutrons from a nearby concrete wall (Gómez-Ros et al., 2018).

In the voxel phantom exercises featuring $^{60}$Co photons and 10 keV neutrons, the instruction given to participants for the location of the point source was to place it '100 cm from the surface of the chest', which could be interpreted differently. In response, the organizers performed small sensitivity analyses to quantify the impact from the ambiguity of this parameter, the outcomes from which were used to imply appropriate 'tolerances' that could be applied to the submitted results (Huet et al., 2022).

The voxel phantom exercise on internal dosimetry was not wisely designed in several respects. The tasks to be solved were too extensive, which also made evaluation challenging and led to delays in feedback to the participants. For electrons and low-energy photons, the source and target organs were sometimes too far apart, which led to very large statistical uncertainties even in the reference solution. Therefore, the degree of deviation between participant and master solutions could not be reliably quantified in some situations.

In the uncertainty exercise, the use of a multi-energy electron source that was similar to the $^{125}$I decay but did not take into account the variability of the actual decay, complicated the understanding of the problem on one hand and, on the other hand, did not favour the analysis of the sensitivity study on the variation of the cross sections. Indeed, the use of monoenergetic electrons would have helped in both aspects.

In the foetal dose exercise, atomic numbers of the elements, mass numbers of the nuclides, and cross-section identifiers had not been fixed for all materials used in the simulations, so participants made their own (different) choices, which caused some of the discrepancies initially noted.

*3.4. Issues with the timeable of the exercises*

Most exercises were planned with a timetable, which in almost all cases proved to be too ambitious and optimistic. This was partly because for many exercises the initial number of participants was lower than was expected and considered adequate for the purpose, so submission deadlines were postponed several times to increase participation after further publicity for the exercises. Further deadline extensions became necessary at the request of the participants.

After an initial analysis of the submitted solutions, participants whose results differed by more than expected from either the reference solution or from most other participants' solutions (as appropriate) were informed on this fact and invited to revise their solutions. Deadlines for the submission of revised results also had to be postponed several times.

As a result, the total duration of the exercises exceeded the typical length of stay of junior researchers at a given institute, making it difficult, if not impossible, to follow-up on abnormal results in some cases.

## 4. Lessons learnt

This section discusses the insights gained from the exercises from the perspective of the organizers.

### 4.1. Problem specification

The participants, as well as the organisers of the exercises, are committed to EURADOS and the intercomparisons in addition to their daily work. Therefore, the topics of the intercomparison exercises must be relevant to the participants' fields of work, and the tasks to be solved should not be overly demanding in terms of setup time. CPU resource requirements may also need to be considered but should generally be less of an issue.

To meet the workload requirements, some of the exercises presented in this Special Issue were designed with simplified idealistic geometric setups and irradiation conditions. Examples include the voxel phantom exercises for monoenergetic point sources, the uncertainty exercise in micro- and nanodosimetry (idealised energy spectrum), and the nanoparticle exercise (simplistic geometry). These simplifications have sometimes raised concerns among reviewers about the usefulness of the respective comparisons but seem justified given the aforementioned time constraints.

Regardless of the complexity of a task, a complete description of the problem to be solved with all relevant information must always be given. For Monte Carlo simulation exercises, this means that the radiation source, simulation geometry and materials must be comprehensively specified. On the other hand, it should generally not be specified exactly how the Monte Carlo simulation or the unfolding are to be carried out. The path to the solution, as well as the tools to be used, must be decided at the discretion of the participant. The participant must determine, for example, whether and which variance reduction techniques can or should be used, whether the transport of secondary charged particles should be simulated or how the thermal neutron transport should be performed.

However, depending on the aim of the exercise, a more detailed specification of intermediate steps or procedures may be advisable. For instance, whenever the performance of codes or their differences is to be assessed, it may be wise also to specify some of the aforementioned aspects of the simulations to ensure that the differences between results from different participants only reflect the differences in the codes that one is interested in.

### 4.2. Reporting of results

The task definition should include very precise instructions for reporting results, and templates should be provided where possible. Providing such a template, where the participants were requested to fill in their results in a pre-defined format, not only helps clarify exactly what output is required from them in each case, but also greatly facilitated the evaluation of the results in the respective exercises. This is especially important when spectral information is to be reported, where a lack of specification of bin division can make synopsis quite cumbersome when different participants use different bin sizes and/or linear and logarithmic equidistant bins.

In addition, asking for redundant information can help to identify potential problems with participants' data. Examples of this were: the voxel phantom exercise for internal dosimetry, where absorbed fractions and specific absorbed fractions were to be reported (differing by only one factor, i.e. organ mass); or the voxel phantom exercise for the X-ray examinations, where the results were to be reported normalised to both kerma free-in-air and kerma area product, which again differ by only one factor. In the case of the nanoparticle exercise, only reporting of results normalised to the number of primary particles was required. If normalisation to the area density of the emitted primary photons from the source had also been reported, the incorrect implementations of the simulation geometry would have been detected much earlier during the exercise.

*4.3. Timing of the exercise*

Regarding the problems encountered with non-responding participants at the revision of results stage, it is planned to set up rules in future exercises to get a more formal commitment from participants. The rules to be established concern deadlines, participation in the feedback loop, and requirements for co-authorship to potential manuscripts (see Supplementary Fig. S1 and Supplementary Fig. S2).

In addition, more timely feedback to the participants might improve their preparedness to disclose details of their computational procedures and improvements. Long feedback intermissions make it difficult for the participants to recall exactly what was done and even what the exercise was about. It should be kept in mind that the organizers, as well as all participants, are performing these exercises alongside their daily duties.

In the reanalysis of the nanoparticle exercise, a set of hierarchical MS Excel templates were used that allowed a fast assessment of the internal consistency of the results reported by participants (in an Excel-template provided to them) as well as a 'live' synopsis via hyperlinks. As an illustration of this approach, Supplementary Fig. S3 to Supplementary Fig. S6 show screenshots of the "Synopsis" worksheets with easy-to-assess graphs and calculated figures of merit (integral quantities normalised so that their expected values are close to unity).

Using these templates to assess the consistency of a participant's results took only a few minutes and required only copying and pasting the results from the Excel templates completed by the participant into the template used for the analysis. In many cases, this enabled feedback in less than an hour. Of course, it was more time-consuming to identify the more sophisticated deviations from the exercise specifications.

Another issue with timing is that calculations with voxel phantoms may require large amounts of time. With some codes, even the visualisation of them can take up to several hours, and the production calculations even weeks. Some codes have ability to skip the geometry check at the beginning of production calculations (e.g. DBCN card in MCNP) and in this way speed up the calculations significantly.

*4.4. Quality assurance of results*

As indicated in Section 3.2, in all exercises some of the participants seemed to have submitted their results without first carrying out adequate quality control of their solutions, e.g. by simple plausibility checks or by comparison with literature data, if available. Approaches such as the one mentioned in Section 4.2 could allow for faster identification of outliers and more timely feedback to participants. This could alleviate some of the problems, such as where there was a lack of response from participants following feedback regarding abnormal results.

However, a significant degradation in the quality of results may persist, as the evolution of many computational tools towards greater ease of use also allows their use without a certain level of expertise, which may still be required for meaningful results. Interaction with some exercise participants who reported unreasonable results revealed a lack of understanding of several fundamental aspects. For example, that the results of calculations are not just numbers, but physical quantities (which have dimensions).

Some participants were also unaware that there are different ways of specifying the categorical variable of a histogram (lower or upper limit of the bin or the bin centre), which can vary between different codes and affect the comparison of results, such as when reported with different definitions of the meaning of the values on the x-axis. This could be countered by requiring both the lower and upper bin limits to be reported.

There were also cases where participants determined ratios with a finer bin size than specified in the task, and then re-binned their results for reporting by averaging the ratios over the larger bins instead of calculating the ratio between the sum of the numerators and the sum of the denominators. Detection of such elementary mistakes requires access to the original results, so their origins were not always immediately apparent to the organizers.

Considering that most exercises will lead to publications in the form of EURADOS reports or journal articles, compliance with the principles of FAIR data (an acronym for findability, accessibility, interoperability and reusability, (Wilkinson et al., 2016)) is also a matter that should be given more emphasis in the future. This is, of course, the responsibility of each participant, but appropriate commitments can be included in the application forms (Supplementary Fig. S1 and Supplementary

Fig. S2). As a minimum requirement, participants should provide documentation on where the following files are stored and backed up:
- data files containing the reported results
- data files with the raw simulation output
- files used for their production (material data or other input files, code, etc.)
- log files and other supplementary output files of the simulations.

This information is essential when participants need to review their work for possible errors. However, in the exercises, delayed responses from participants were sometimes explained by difficulties in finding data, uncertainty about which version of a code was used, and similar such problems. Therefore, it may be useful for the organisers to collect this information - or even the files containing the metadata of the simulations - as part of the reporting of the results. This might be the case especially for participants who are inexperienced users of simulation codes, who may not be aware that the log files etc. generated by their codes contain information that complements their simulation results and is important for their quality.

It was not uncommon for files of original simulation results, which were shared by participants with the organisers during the feedback loops, to contain only columns of numbers. A header indicating which quantities are listed and which units have been used, and ideally also containing information on the code used and its version, as well as the date when the file was written, would be minimum requirements for the useability of these data files. In addition, the aforementioned auxiliary information is also needed.

## 5. Conclusions

Beyond doubt, the reported EURADOS exercises are beneficial to the field of computational dosimetry. They directly contribute to the training of the participants by improving their computational procedures through feedback with the task organisers. They lead also to the availability of representative dose values for various exposure conditions that may aid future novice users in the quality assurance of their methods. In addition, they also provide a snapshot of how well (or otherwise) the computational techniques are being applied within the community in general, and how well some of the concepts recommended by organizations such as ICRP are understood; the observed difficulties in correctly defining and evaluating effective doses (Eakins et al., 2021; Huet et al., 2022), or in determining bone marrow doses (Zankl et al., 2021a), are clear examples of the latter.

The obvious question of what could be done better in future exercises has been partly addressed in Section 4. A general answer to this question is not easy, since it depends on the objective of the exercises. The question will therefore continue to be the subject of discussion within EURADOS Working Group 6 when new exercises are prepared.

To avoid participants wasting their time on the tasks of an exercise in cases where they are prone to give incorrect results, due to a wrong idea of the task or ignorance of the dosimetric quantities to be determined, several modifications of the exercises can be considered. One could be to define explicitly the dosimetric quantities and normalisation quantities to be used. Another possibility would be to include in the definition of the task a list of checks to be made by the participants on the results, or even provide them with templates like those used in the reanalysis of the nanoparticle exercise.

Such changes to the exercises would mitigate the risk of potential errors by the participants. However, while this closer guidance may lead to improved results, this better agreement will only reflect the participants' capability to follow detailed instructions and not their actual state of expertise or their performance in real-world applications. Moreover, concepts like effective dose are widely used and already well-defined elsewhere; arguably, it should not therefore be the role of EURADOS WG6, which focusses on computational dosimetry, to coach radiation professionals on the basic concepts that underpin radiological protection.

Most of the exercises were the types of task that the participants may be confronted with in their professional activities or in research. Generally, they then would have to perform the simulation or unfolding, and calculate the dosimetric quantities of interest, without specific guidance. A supervisor of an early-stage researcher, or a reviewer of potential papers arising from such work, can be a source of feedback that hopefully reveals major non-plausibilities in the results, if any. However, as discussed by (Rabus et al., 2021c), these potential quality filters often seem to have failed for simulation studies on nanoparticles so far.

A better option could therefore be to create a questionnaire to assess the knowledge of the participants, and then give more detailed instructions to the less experienced participants. One could also start the exercise with a webinar explaining what is expected, what participants should do, and explaining the importance of quality assurance. A recording of the webinar could also be made available on the EURADOS website so that participants who join the exercise later can refer to it.

Another potential improvement for some of the exercises, if repeated, could be to include several reporting steps in the exercise. For example, in the voxel phantom exercise for the X-ray examinations, participants could first need to report their results for the position of the radiation source in relation to the phantom; they could then get feedback if it is outside the uncertainty band of the reference value and be invited to report a revised value. On request they could also get the correct positions as feedback and run their simulations for these. Alternatively, such future exercises could have a first step with as few specifications as possible, a second with plausibility checks to be performed by the participants, and a third with feedback on the submitted results and a request for revision.

## Acknowledgements

The support of all members of EURADOS WG 6 who contributed to the organization of the exercises and the efforts of the participants to solve the tasks are gratefully acknowledged. The authors also thank the Editor-in-Chief of Radiation Measurements, Adrie de Bos, for making this special issue possible, and the EURADOS Council for providing the necessary funding.

**Supplement**

*Supplementary Figures*

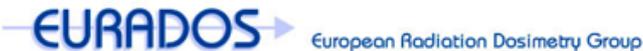

**Supplementary Fig. S1:** Sample application form for participation in a EURADOS WG 6 intercomparison exercise, with set rules in data handling and conduct by participants, for the case of an exercise with restricted authorship in publications on the results.

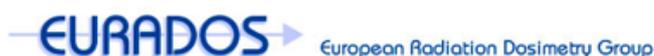

**Supplementary Fig. S2:** Sample application form for participation in a EURADOS WG 6 intercomparison exercise, with set rules in data handling and conduct by participants, for the case of an exercise with co-authorship of participants.

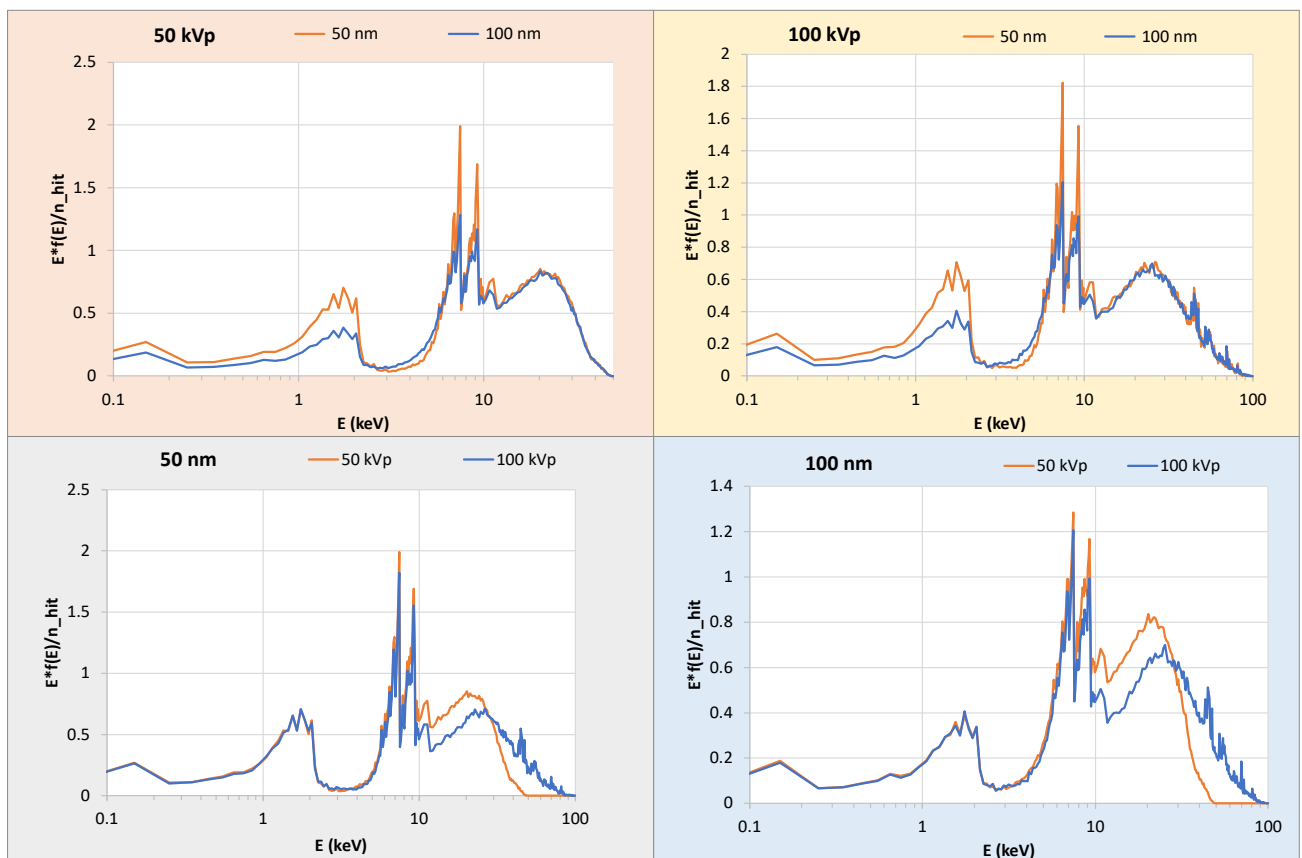

**Supplementary Fig. S3:** Screenshot of the section with the diagrams in the "Synopsis" worksheet of the Microsoft Excel template used in the nanoparticle exercise to test the internal consistency of the results reported by a participant for the energy spectrum of electrons emitted from the gold nanoparticle for the four combinations of nanoparticle size and X-ray spectrum. The plots show comparisons of the energy distributions of the number of emitted electrons (normalised to the average number of photon interactions in the gold nanoparticle for the simulation geometry, as defined in the exercise). The two plots in the upper row compare different nanoparticle sizes for the same radiation quality, where similar values are expected for high-energy electrons and higher frequencies for low-energy electrons for the smaller nanoparticle. The bottom row shows the comparison for the same nanoparticle size and different energy spectra, where one expects similar values for low electron energies because the range of these electrons is smaller than the size of the nanoparticle. The data shown have been calculated from the participants data by rebinning and normalization to the expected number of photon interactions in the nanoparticle. They are plotted in "microdosimetry style" such that the area under the curves is proportional to the number of electrons emitted per photon interaction in the respective energy interval.

The data for emitted electrons (and the corresponding data for energy deposition shown in Supplementary Supplementary Fig. **S4**) correspond to one of the cases where the participant correctly applied the required geometry. Examples of how different the respective graphs look for the case of deviating geometry can be found in (Rabus et al., 2021c).

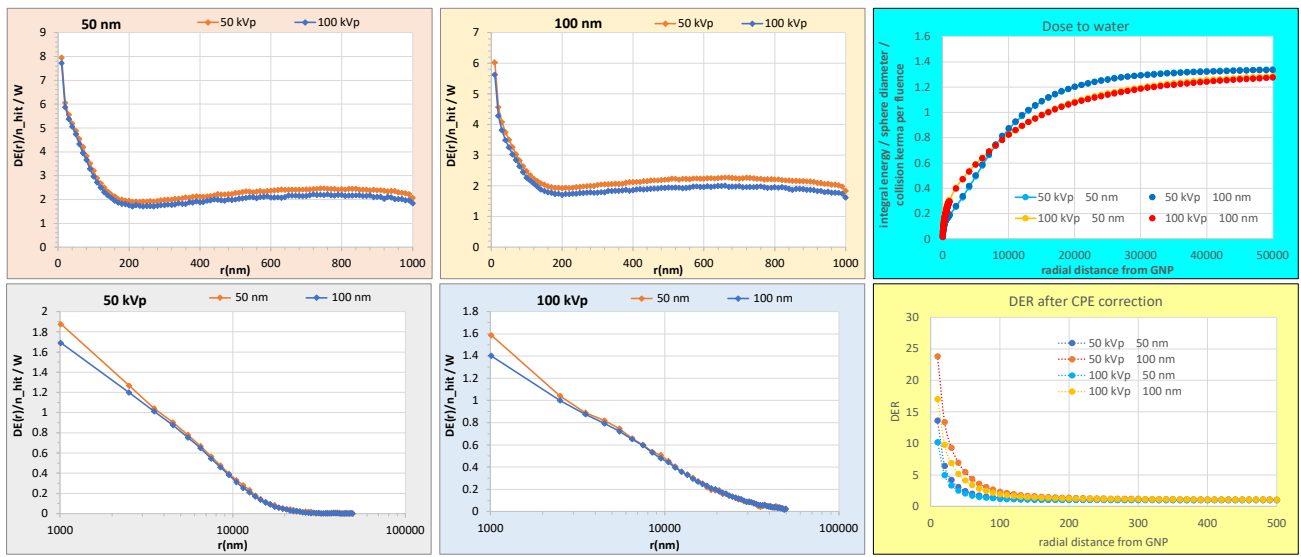

**Supplementary Fig. S4:** Screenshot of the section with the diagrams in the "Synopsis" worksheet of the Microsoft Excel template used in the nanoparticle exercise to test the internal consistency of the results reported by a participant for the energy deposition in spherical shells around a gold nanoparticle. The four plots in the left and middle columns show comparisons between the results for the different cases (two radiation qualities, two nanoparticle diameters) after conversion to easily understood quantities (number of ionizations in the spherical shells around the nanoparticle per photon interaction in the nanoparticle). The diagram on the upper right (blue frame) shows a comparison of all four data sets of the absorbed dose convergence for water as the size of the scoring region increases. The bottom right plot (yellow frame) shows the results for the quantity of interest (dose enhancement ratio) for the four cases studied. For details see (Rabus et al., 2021b, 2021c).

**Supplementary Fig. S5:** Screenshot of the data section in the "Synopsis" worksheet of the Microsoft Excel template used within the nanoparticle exercise to test the internal consistency of the results reported by a participant for the energy deposition in spherical shells around a gold nanoparticle. The values in lines 22 ff. are calculated from the participant's data by applying the correction factors entered by the user in the white fields of lines 7 and 8. The numbers in rows 17 and 19 show figures of merit calculated from the data. For details see (Rabus et al., 2021b, 2021c).

| | 50 kVp, 50 nm | | 50 kVp, 100 nm | | 100 kVp, 50 nm | | 100 kVp, 100 nm | |
|---|---|---|---|---|---|---|---|---|
| **Data set** | | | | | | | | |
| Integral / mean energy | 0.90 | | 0.85 | | 0.92 | | 0.88 | |
| above / bin width | 0.18 | | 0.17 | | 0.18 | | 0.18 | |
| | | | | | | | | |
| Correction factor applied | 1/5 | 0.2000 | 1/5 | 0.2000 | 1.00 | 1.0000 | 1/5 | 0.2000 |
| | | | | | | | | |
| | E (eV) | f(E) | E (eV) | f(E) | E (eV) | f(E) | E (eV) | f(E) |
| | 2.5 | 0 | 2.5 | 0 | 2.5 | 0 | 2.50E+00 | 0.00E+00 |
| | 7.5 | 0 | 7.5 | 0 | 7.5 | 0 | 7.50E+00 | 0.00E+00 |
| | 12.5 | 1.71E-06 | 12.5 | 2.07E-06 | 12.5 | 8.66E-07 | 1.25E+01 | 1.22E-06 |
| | 17.5 | 1.83E-06 | 17.5 | 2.37E-06 | 17.5 | 8.76E-07 | 1.75E+01 | 1.15E-06 |
| | 22.5 | 2.04E-06 | 22.5 | 2.69E-06 | 22.5 | 9.94E-07 | 2.25E+01 | 1.34E-06 |
| | 27.5 | 2.00E-06 | 27.5 | 2.67E-06 | 27.5 | 9.26E-07 | 2.75E+01 | 1.22E-06 |
| | 32.5 | 2.01E-06 | 32.5 | 2.51E-06 | 32.5 | 9.82E-07 | 3.25E+01 | 1.34E-06 |
| | 37.5 | 2.54E-06 | 37.5 | 3.12E-06 | 37.5 | 1.23E-06 | 3.75E+01 | 1.67E-06 |
| | 42.5 | 2.16E-06 | 42.5 | 2.97E-06 | 42.5 | 1.13E-06 | 4.25E+01 | 1.39E-06 |
| | 47.5 | 1.77E-06 | 47.5 | 2.53E-06 | 47.5 | 9.18E-07 | 4.75E+01 | 1.20E-06 |
| | 52.5 | 1.81E-06 | 52.5 | 2.48E-06 | 52.5 | 8.00E-07 | 5.25E+01 | 1.19E-06 |
| | 57.5 | 2.12E-06 | 57.5 | 2.90E-06 | 57.5 | 1.03E-06 | 5.75E+01 | 1.39E-06 |
| | 62.5 | 2.06E-06 | 62.5 | 3.03E-06 | 62.5 | 1.04E-06 | 6.25E+01 | 1.42E-06 |
| | 67.5 | 2.00E-06 | 67.5 | 2.89E-06 | 67.5 | 9.48E-07 | 6.75E+01 | 1.51E-06 |
| | 72.5 | 3.64E-06 | 72.5 | 4.81E-06 | 72.5 | 1.87E-06 | 7.25E+01 | 2.52E-06 |

Header of table (yellow banner): **Correction: Normalization to bin width where needed**

**Supplementary Fig. S6:** Screenshot of the data section in the "Synopsis" worksheet of the Microsoft Excel template used within the nanoparticle exercise to test the internal consistency of the results reported by a participant for the energy spectrum of electrons emitted from the gold nanoparticle. The values in lines 18 ff. are calculated from the participant's data by applying the correction factors entered by the user in line 15. The numbers in row 12 show figures of merit calculated from the data. For details see (Rabus et al., 2021b, 2021c).